\documentstyle[aps,prl,epsf]{revtex} %twocolums

\begin{document}
\twocolumn[{  %twocolums
\draft        %twocolums
\title
{\bf 
%%%%%%%%%%%%%%%%%%%%%%%%%%%%%%%%%%%%%%%%%%%%%%%%%%%%%%%%%%%%%%%%%%%%%%%%%%%%%%%
%%%%%%%%%%%%%%%%%%%%%%%%%%%%%%%%%%%%%%%%%%%%%%%%%%%%%%%%%%%%%%%%%%%%%%%%%%%%%%%
%                    The Truth About Earthquakes. Part III:
                           Time-Decreasing Hazard and
                   Increasing Time until the Next Earthquake
%%%%%%%%%%%%%%%%%%%%%%%%%%%%%%%%%%%%%%%%%%%%%%%%%%%%%%%%%%%%%%%%%%%%%%%%%%%%%%%
%%%%%%%%%%%%%%%%%%%%%%%%%%%%%%%%%%%%%%%%%%%%%%%%%%%%%%%%%%%%%%%%%%%%%%%%%%%%%%%
}
\author
{
%%%%%%%%%%%%%%%%%%%%%%%%%%%%%%%%%%%%%%%%%%%%%%%%%%%%%%%%%%%%%%%%%%%%%%%%%%%%%%%
%Santi Monguitoh %$^1$\cite{underg},
%and
\'Alvaro Corral%
%\footnote{
%%Grup de F\'\i sica Estad\'\i stica,
%Departament de F\'\i sica, Facultat de Ci\`encies, 
%Universitat Aut\`onoma de Barcelona,
%%Edifici Cc, 
%E-08193 Bellaterra, Barcelona, Spain 
%}
%$^{1,2}$%$^{\dagger,}$ 
\cite{email} 
%%%%%%%%%%%%%%%%%%%%%%%%%%%%%%%%%%%%%%%%%%%%%%%%%%%%%%%%%%%%%%%%%%%%%%%%%%%%%%%
}
\address{
%%______________________________________________________________________________
%%%$^{\dagger}$%
%%$^1$
%%The Niels Bohr Institute, University of Copenhagen,       
%%Blegdamsvej 17,  DK-2100  Copenhagen \O,   Denmark \\
%%$^2$
%Grup de F\'\i sica Estad\'\i stica,
Departament de F\'\i sica, Facultat de Ci\`encies, 
Universitat Aut\`onoma de Barcelona,
%Edifici Cc, 
E-08193 Bellaterra, Barcelona, Spain 
%%-------------------------------------------------------------------------------
}
%
%_______________________________________________________________________________
%%\date{Accepted in PRE 15 July 2002; published ???}
\date{\today}
%-------------------------------------------------------------------------------

\maketitle %\parskip 2ex
\widetext  %twocolums
\begin{abstract}
\leftskip 54.8 pt %twocolums
\rightskip 54.8 pt %twocolums 
%____________________________ ABSTRACT ________________________________________
The existence of a slowly %%decaying %%power-law distribution (OJO!!) 
always-decreasing probability density for the recurrence times
of earthquakes implies that the occurrence of an event at a given instant
becomes more unlikely as time since the previous event increases.
Consequently, the expected waiting time to the next earthquake
increases with the elapsed time, 
that is, the event moves away to the future fast. %in time.
We have found direct empirical evidence of this counterintuitive behavior
in two worldwide catalogs as well as in diverse local catalogs.
Furthermore, the phenomenon %, which is valid for all times,
can be well described by universal scaling functions.
%%This counterintuitive behavior, which is valid for all times,
%%has been directly measured in two worldwide catalogs
%%as well as in diverse local catalogs, and %%we find that these behaviors 
%%can be well described by universal scaling functions.
%-------------------------------------------------------------------------------
\end{abstract}
\leftskip 54.8 pt %twocolums
}]   %twocolums

\narrowtext
%
%\setcounter{page}{1} %twocolums
%_______________________________________________________________________________
%\markright{\bf \today} %twocolums
%-------------------------------------------------------------------------------
%\thispagestyle{myheadings} %twocolums
%\pagestyle{myheadings} %twocolums
%
%\newpage   %twocolums % este va al reves

%%%%%%%%%%%%%%%%%%%%%%%%%%%%%%%%%%%%%%%%%%%%%%%%%%%%%%%%%%%%%%%%%%%
%\section{ Introduction.}
%%%%%%%%%%%%%%%%%%%%%%%%%%%%%%%%%%%%%%%%%%%%%%%%%%%%%%%%%%%%%%%%%%%

Many probability distributions have been 
%tested to evaluate their likelihood (as) to describe 
proposed to account for
the recurrence time of earthquakes \cite{Smalley,Sornette97,Wang,Ellsworth}, 
which is the time interval between successive
earthquakes in a certain region.
When aftershocks and mainshocks are considered together 
%
%as essentially the same type of event,
%
%as essentially the same phenomenon,
%
%as essentially part of a unique phenomenon,
%
%as a part of essentially one unique phenomenon,
%
%as essentially part of a unique process,
%
%as a part of essentially the same process,
%
as a part of essentially one unique process \cite{Bak02,Christensen},
we have determined that 
a universal scaling law describes the probability
density $D(\tau)$ of the recurrence time.
In this way, for events 
of magnitude $M$ above a certain threshold $M_c$
in a given spatial area 
(whose limits do not need to depend on the tectonic background),
$D(\tau)$ scales with the rate of seismic activity $R$ in the area
as
$$
D(\tau)=R f(R \tau),
$$
where $R$ 
is defined as the mean number of earthquakes (with $M \ge M_c$) per unit time
and $f$ is a universal scaling function.
This scaling %reveals the self-similarity 
is in fact the hallmark of the self-similarity 
of seismicity in the space-time-magnitude domain.
Among several possible scaling functions, 
the best fit is obtained 
%The scaling function can be represented 
from a (truncated) gamma distribution,
$$
    f(\theta) = \frac{C}{a \Gamma(\gamma)}
           \left(\frac{a}{\theta} \right)^{1-\gamma}
           e^{-\theta/a},
$$
with 
$\gamma$ the shape parameter, $a$ the scale parameter, 
$C$ a correction to normalization
(due to the truncation of the distribution close to zero),
and $\Gamma(\gamma)$ the gamma function.
When $\gamma <1$,
$f$ turns out to be a decreasing power law,
accelerated by an exponential factor in the long-time limit.

%%The parameters are determined by a least-square fit to be
We analyze %%rescaled recurrence times for 
many regions and $M_c-$values
from two worldwide catalogs (the NEIC-PDE and  the 
Significant Earthquake Database from NOAA at NEIC \cite{NEIC}) 
and from several regional 
catalogs (Southern California, Japan, 
New Zealand, New Madrid (USA),
the Iberian Peninsula, and the British Islands \cite{catalogs}).
Each analyzed region is delimited by two meridians and two parallels
\cite{except},
with linear size spanning from $0.16^\circ$ (about 18 km)
to the whole globe ($ \sim 20 \cdot 10^3$ km), covering 
a large variety of tectonic environments, whereas 
the considered magnitudes range from larger than 1.5
to larger than 7.5
(this is about a factor $10^9$ in the minimum released energy).

Except for a $360^\circ \times 180^\circ-$region, 
which covers the entire globe, the rest of the regions
are defined by a window of $L$ degrees in longitude 
and $L$ degrees in latitude.
The coordinates $(x,y)$ of the west-south corner of these regions
can be obtained from the vector $(k_x, k_y)$ at Fig. 1's labels as 
$x = x_{min} + k_x L$, $y = y_{min} + k_y L$, with 
$(x_{min}, y_{min}) = (-180^\circ, -90^\circ)$, $(-123^\circ, 30^\circ)$, 
$(127^\circ, 27^\circ)$, $(160^\circ, -60^\circ)$, $(-91^\circ, 35^\circ)$, 
$(-20^\circ, 30^\circ)$, and $(-10^\circ, 45^\circ)$ 
for the worldwide, Southern California, Japan, New Zealand, New Madrid, 
Iberian Peninsula, and British Island catalogs,
respectively. 
The periods of study are (in years A.D. including the last one) 
1973-2002 for the NEIC catalog, 1897-1970 for the NOAA one,
1988-1991, 1995-1998, and 1984-2001 for Southern California
(denoted as SC88, SC95, and SC84), and 1995-1998, 1996-2001, 1975-2002,
1993-1997, and 1991-2001, for the rest of catalogs
(in the same order as before).  
The regions and times of observation are selected in order
%to include periods of stationary seismic activity,
that a period of stationary seismic activity is included,
this means that aftershock sequences should not have too much weight
in the seismicity of the region. 
When this is not the case (i.e., for very large aftershock activity)
our analysis is still valid, but the scaling with the mean rate 
has to be replaced by a scaling with the instantaneous rate.

A maximum-likelihood estimation of the parameters 
using the rescaled recurrence time $R \tau$
for all the regions and $M_c$'s studied
gives $\gamma= 0.74 \pm 0.05$,  
and $a = 1.23 \pm 0.15$, %and %$C = 1.05 \pm 0.10$,
which yields a coefficient of variation $CV \simeq 1.2$. 
The constant $C$ is determined from the normalization condition
given the minimum value for which the gamma distribution holds;
for $\theta > 0.05$, $C=1.10 \pm 0.10$ (see Fig. 1 (a)).

%Figure 1A shows 
The results of the fit are shown in Fig. 1 (a) using the survivor function,
which is defined as 
$S(\tau) \equiv \mbox{Prob} [\tau' > \tau]=\int_\tau^\infty D(\tau') d \tau'$
(where $\tau'$ is a generic label for the recurrence time,
while $\tau $ refers to a particular value of the same quantity).
It is straightforward to obtain that, in our case, $S(\tau)$ should also verify a
scaling relation, $S(\tau)=G(R \tau)$, with 
$G(\theta)=C Q_\gamma(\theta /a)$, and 
$Q_\gamma(\theta/a)$ the complement of the incomplete gamma function
\cite{Abramowitz,Press}.
%Figure 1A shows the scaling with $R$ of recurrence-time survivor functions
%The data collapse in the plot is in total agreement with these equations.
The total agreement between these equations and the measured distributions
is clear from the data collapse and the fitting curve in the plot, 
for intermediate and long recurrence times.
The accuracy of the scaling law and the gamma fit is
%clear from the plot
%for intermediate and long recurrence times, about
%$\tau > 0.01 / R$,
guaranteed %provided that 
as the seismic activity is stationary 
in this range of recurrence times.
%When large variations of activity are present,
%due to the triggering of aftershock sequences, 
%both the scaling and the gamma distribution are
%still valid, if the scaling is undertaken with
%the time-varying rate, rather than with its mean value.
On the contrary,
short times are usually not free of disturbances of the stationariness,
due to the triggering of aftershock sequences, which destroy
the universal scaling behavior.
%is not valid for short times
%(at least if one looks at the process in the mean-rate scale).
%Therefore, %to get a good scaling of the data 
%it is more convenient
In order to treat all the distributions in the same way,
we calculate the rate $R$ %with some degree of homogeneity...
only for events in the scaling region, i.e., 
short recurrence times are not considered in the rate.
%(at least if one looks at the process in the mean-rate scale).
%due to their non-universal behavior...
%Nevertheless, the universal scaling behavior can be recovered if
%the scaling is undertaken with
%the time-varying rate, rather than with its mean value.

%, which is obtained analyzing
%%many regions and $M_c$, 
%%using 
%time intervals from the NEIC-PDE worldwide catalog and from regional 
%catalogs for Southern California, Japan, 
%New Zealand, New Madrid (USA),
%the Iberian Peninsula, and the British Islands
%CITAR!!!.  
%The linear size of the analyzed regions spans from $0.16^\circ$ (about 18 km)
%to the whole globe ($ \sim 20 \cdot 10^3$ km), covering 
%a large variety of tectonic environments,
%whereas the considered magnitudes range from larger than 1.5
%to larger than 7.5
%(this is about a factor $10^9$ in the minimum released energy). 

The knowledge of the probability distribution 
of the recurrence times allows
one to answer two important questions about the 
temporal occurrence of earthquakes.
First, for a certain region and for $M \ge M_c$,
we can study the probability per unit time of 
an immediate earthquake given that there has been a period 
$\tau$ without activity, using the hazard rate \cite{Kalbfleisch},
$$
\lambda(\tau) \equiv 
       \frac{ \mbox{Prob} [\tau < \tau' \le \tau + d \tau \, | \, \tau' > \tau ]}{d \tau}
%       =\frac{D(\tau)}{\int^{\infty}_{\tau}D(\tau ') d \tau '},
       =\frac{D(\tau)}{S(\tau)},
$$
where 
%$\tau'$ is a generic label for the recurrence time,
%while $\tau $ refers to a particular value of the same quantity, and
the symbol ``$|$'' accounts for conditional probability.
From the previous formulas we get that $\lambda(\tau)$ 
scales as $\lambda(\tau)=R h(R \tau)$, with
%It is easy to obtain that, in our case, $\lambda(\tau)$ should also verify a
%scaling relation, $\lambda(\tau)=R h(R \tau)$, with
$$
h(\theta)=\frac{1}{a \Gamma(\gamma)}  \left(\frac{a}{\theta} \right)^{1-\gamma}
           \frac{e^{-\theta/a}}{Q_{\gamma}(\theta/a)}.
$$
%where $Q_\gamma(\theta/a)$ is the complement of the incomplete gamma function
%{\it (8,9)}.
%The hazard rate $\lambda(\tau)$ 
This function
turns out to be monotonically decreasing, 
tending as a power law to the value $1/a$ as $\theta \rightarrow \infty$.
So, contrary to common belief, the hazard does not increase with
the elapsed time since the last earthquake, but just the opposite;
this is precisely the more direct characterization 
of long-term clustering \cite{Kagan91}.

Also, one can wonder about the expected time till the next earthquake,
given that a period $\tau_0$ without earthquakes 
(in the spatial area and range of magnitudes considered) has elapsed,
$$
\epsilon(\tau_0) \equiv
E[\tau-\tau_0 \, | \, \tau > \tau_0]=
\frac{1}{S(\tau_0)}
{\int^{\infty}_{\tau_0} (\tau-\tau_0) D(\tau) d \tau}.
%                              {\int_{\tau_0}^{\infty} D(\tau) d \tau}.
$$
This function can be referred to as the expected residual recurrence time 
\cite{Kalbfleisch}
and again we find a scaling form for it, which is 
$\epsilon(\tau_0)=e(R \tau_0) / R$, with the scaling function 
$$
e(\theta)=a \gamma \frac{Q_{\gamma+1}(\theta/a)}{Q_{\gamma}(\theta/a)} - \theta.
$$
This is an increasing function of $\theta$, which reaches an asymptotic value
$e(\theta) \rightarrow a$.
Therefore, the residual time until the next earthquake should grow with the elapsed
time since the last one.
Notice the counterintuitive behavior that this represents:
if we decompose the recurrence time $\tau$ as $\tau=\tau_0 + \tau_f$,
with $\tau_f$ the residual time to the next earthquake, the increase of $\tau_0$
implies the increase of the mean value of $\tau_f$, but the mean value
of $\tau$ is kept fixed.
%This is sometimes referred to as the waiting-time paradox.
In fact, this is just a more dramatic version of the classical 
waiting-time paradox \cite{Feller,Schroeder}.

For the particular case of earthquakes
this is even more paradoxical, 
since one would say that the longer the time one has been waiting
for an earthquake, the closer it will be,
due to the fact that as time passes 
stress increases on the corresponding fault
and the next earthquake becomes more likely.
(Nevertheless, one needs to distinguish between 
earthquakes on a given fault and earthquakes 
over a certain area.)
The question was originally put forward by Davis {\it et al.} \cite{Davis},
who pointed out that if a lognormal distribution is a priori assumed
for the recurrence times, the expected residual time increases
with the waiting time.
(However, the increase here was associated to the update of the
distribution parameters as the time since the last earthquake,
which was taken into account in the estimation, increased;
we deal with a different concept of increasing residual time.)
Sornette and Knopoff \cite{Sornette97} showed that 
the increase (or decrease) depends completely
on the election of the distribution, and studied 
the properties of a number of them.
We are going to see that the observational data provide 
direct evidence against the simple picture of the next earthquake
approaching in time.

Indeed, our mathematical predictions can be contrasted with the catalogs;
both the hazard rate and the expected 
%time till the next earthquake at a given elapsed time 
residual recurrence time
can be directly measured
with no assumption about their functional form.
Following their definitions, these quantities are estimated as
$$
  \lambda(\tau) = \frac{n(\tau,\tau+\Delta \tau) }
                {n(\tau,\infty)
                  \Delta \tau },
%$$
%$$
%\, 
\hspace{0.5cm}
  \epsilon(\tau_0) = \frac{\sum_{i 
                | \tau_i > \tau_0} 
 (\tau_i-\tau_0)}
               {n(\tau_0,\infty)},
$$
where $n(\tau_1,\tau_2)$ denotes the number of quakes with recurrence time 
in the range $(\tau_1,\tau_2)$ and the sum in $\epsilon(\tau_0)$ is computed only for earthquakes $i$
such that $\tau_i > \tau_0$ (and of course $M \ge M_c$).
From the results displayed in Figs. 1 (b) and 1 (c)
it is apparent that the hazard rate decreases with time whereas the expected
residual recurrence time increases, as we predicted.
Moreover, both quantities are well approximated  by the proposed universal
scaling functions.

%A typical behavior 
%Regarding 
On the other hand,
the part of the recurrence-time distribution 
that accounts for short times displays a typical behavior 
$f(\theta)=K_1/\theta^{1-\alpha}$,
%A rough upper limit for $D(\tau)$ in this case is $K_1/\tau^{1+\alpha}$,
and the corresponding functions for the survivor function, hazard rate,
and expected residual return time turn out to be:
$$
G(\theta)=K_1(K_2-\theta^\alpha/\alpha),
\hspace{0.5cm}
h(\theta)=\frac{1}{K_2 \theta^{1-\alpha}-\theta/\alpha},
%{\theta/\alpha+K_2 \theta^{1+\alpha}},
$$
$$
%\hspace{0.5cm}
e(\theta)=\frac{K_3 - \theta^{1+\alpha} / (1+\alpha)}
               {K_2 - \theta^\alpha/\alpha}-\theta,
$$
where the constants $K_2$ and $K_3$ depend on the rest of the distribution.
An example for these functions with $\alpha \simeq 0.2$ is also represented in Fig. 1, 
showing the appropriate decreasing or increasing tendency 
in each case.

%-------------------------- ELIMINADO!!!!!--------------------------
%
%Finally, it is worth mentioning that 
%in all the scaling laws the rate $R$
%can be replaced by its dependence with the magnitude, $R \sim 10^{A-b M}$
%{\it (5,6)};
%nevertheless, one has to have in mind that, mainly the parameter $A$, 
%but also $b$,
%are nonuniversal, in the sense that they present regional variations.
%
%-------------------------------------------------------------------

For the sake of concreteness, let us consider 
worldwide earthquakes with $M \ge 7.5$, 
%whose rate is $R=6$ per year, roughly.
which occur at a rate $R=6$ per year, roughly.
In the days immediately after one event of this type, the expected
time to the next one (anywhere in the world) is about 2 months 
(for $\tau=6$ days, we have $R\tau=0.1$, 
and $e(0.1)\simeq 1$, see Fig. 1 (c)).
If after 2 months the quake has not come, 
the expected residual time not only does not decrease but
increases to 2.2 months ($e(1) \simeq 1.1$,
this would lead to $4.2$ months between both events),
and if the elapsed time reached 1 year 
(which is unlikely but not impossible), 
the expected waiting time would further increase 
to 2.4 months ($e(6) \simeq 1.2$).
In the same way, the hazard rate would drop from 0.7 
to 0.5 and to 0.4 month$^{-1}$
($h(\theta) \simeq 1.4, 1$, and $0.85$), respectively.
The same process is reproduced at all magnitude and
spatial scales in a self-similar manner.
An intuitive explanation of this phenomenon is that
when the elapsed time since the last earthquake is large, 
the system enters into a long ``drought period"
in which the recurrence time is likely to be very large.
Notice however that there is no fundamental % abrupt
difference between these drought periods and the rest of recurrence times,
since all of them are governed by the same smooth distribution.

The universality of this behavior demands further explanation;
nevertheless, it suggests the existence of a simple mechanism 
in which, as time passes, the variable that triggers rupture 
runs away from the rupture threshold (on average).
%but eventually it will return to the threshold.
The ``excursions" of this variable would keep the memory 
of the last event stored in the system 
up to very long times to generate the {\it negative aging} observed.

The considerations reported here should be at the core
of any research regarding earthquake-occurrence modeling \cite{Ogata,Helmstetter}
and predictability \cite{Main,Geller,nature,Parsons}.
Finally,
in order to account for the self-similarity of these processes,
the concept of self-organized criticality 
provides the most appealing framework up to now \cite{Bak96}.

This author has benefited a lot from the original perspectives
and deep insights of the late Per Bak.
%who also had the idea to sell insurances in places
%where people is long waiting for an earthquake.
%It should work.
The author also thanks M. Bogu\~n\'a, D. Sornette's criticisms,
the Ram\'on y Cajal program of the Spanish MCyT, 
and all the people at the Statistical
Physics Group of the Universitat Aut\`onoma de Barcelona, 
as well as those institutions that have
made their catalogs available on the Internet.

%-------------------------- ELIMINADO!!!!!--------------------------
%
%As an illustration, for the whole Southern-California area
%the stationary rate for $M \ge 5.5$ events is about
%1 event each year.
%An extrapolation of $\epsilon(\tau)$ to this case 
%implies that if the last event in this range took place
%0.1 years ago, we will have to wait somewhat less than one
%year to the next event; however, if the last event would have
%taken place 10 years ago, the expected waiting time 
%would increase to 1.2 years.
%Notice that this increase is not a consequence of the large
%value of the elapsed time, which does not enter into the calculation.
%%% valores de la funcion, e(0.1) < 1, e(1)=1.1, e(10)=1.2
%
%----------------------------------------------------------------

%The considerations reported here should be at the core
%of any research regarding earthquake-occurrence modeling {\it (14,15)}
%and predictability {\it (16-19)}.
%Additionally, the universality of the
%time-decreasing hazard and time-increasing expected residual time
%calls for a unified framework for earthquake description,
%where the memory of the last event is stored in the system 
%up to very long times to generate the {\it negative aging} observed.
%So far, self-organized criticality provides the most appealing picture
%for all these phenomena {\it (20)}.

%\end{document}

%\vspace{1cm}
%{\bf Figure Caption}
%{\bf Fig. 1.} 

\newpage

%\thispagestyle{empty}
%
%########################################################################
\begin{figure}
\epsfxsize=3.5truein %\hskip 1.2truein %2.75 %3
\epsffile{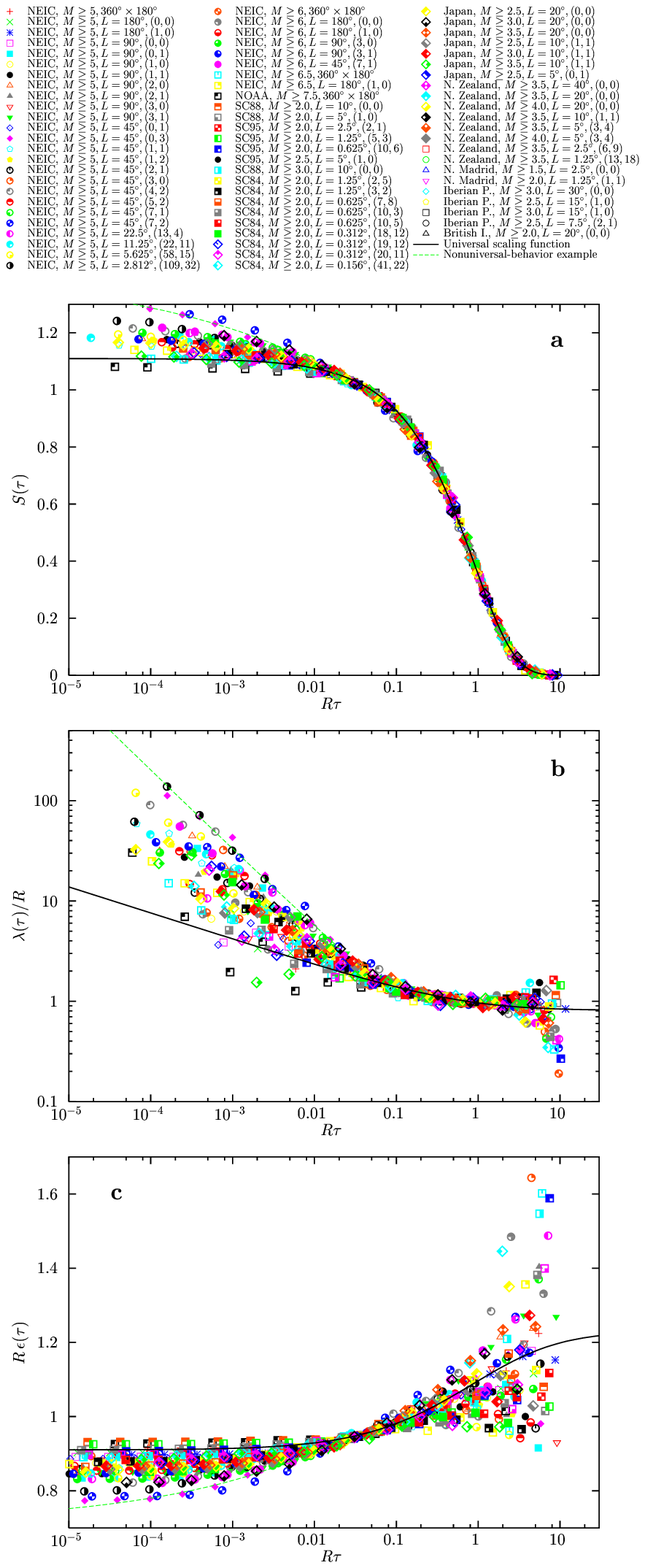}
\caption{
Scaling plots of the probability distributions, hazard rate functions,
and expected residual recurrence times for all the catalogs
analyzed.
The values of the occurrence rates $R$ are broadly distributed,
ranging roughly from 6 year$^{-1}$ % 1.8e-7 Hz NOAA
to 1 hour$^{-1}$, % 3.2e-4 Hz Japan L=20, Mc=2.5
so one unit in the horizontal axis can represent from 1 hour to 2 months.
%in the cases studied here.
The universal scaling functions fitting the data 
are the ones proposed in the text,
with the parameters obtained from the maximum-likelihood estimation;
an example of fit outside the scaling region is also shown.
{(a)} 
Rescaled survivor functions.
The distributions are normalized for $R \tau \ge 0.05$,
therefore, the left part of the distributions does not 
represent a probability;
nevertheless, we have considered interesting to
show it to illustrate the nonuniversal behavior.
Times shorter than two minutes are not shown.
{(b)} 
Rescaled hazard rates. 
The errors at long times are large,
due to poor statistics.
{(c)} 
Rescaled expected residual recurrence times.
Only mean values calculated with 3 or more data are displayed.
At long times the errors increase even further in this case, 
as $\epsilon(\tau)$ is the difference of two large numbers;
nevertheless, the gaps between the points and the function 
are compatible with the error bars (not shown).
}
\end{figure}
%########################################################################
%

\end{document}